\documentclass[aps,pra,twocolumn,superscriptaddress,showpacs,floatfix]{revtex4}
\usepackage{graphicx}
\usepackage{color}
\usepackage{bm}
\usepackage{amsmath}

\newcommand{\fig}[1]{Fig.~\ref{#1}}
\newcommand{\be}[1]{\begin{equation}\label{#1}}
\newcommand{\ee}{\end{equation}}
\renewcommand{\vec}[1]{{\boldsymbol #1}}

\begin{document}

\title{Intensity dependence of ionization mechanisms for infrared frequencies in the strong field double ionization of diatomic molecules}

\author{Agapi Emmanouilidou}
\affiliation{Chemistry Department, University of Massachusetts at Amherst, Amherst, Massachusetts, 01003} \date{\today}

\date{\today}

\begin{abstract}
Using a three-dimensional quasiclassical technique we explore the double ionization pathways of a diatomic molecule driven by an intense infrared (800nm) ultrashort laser pulse.
For intensities corresponding to the tunneling regime, we find that the three main ionization mechanisms have distinct traces when considering the sum of the momenta parallel to the laser field
 as a function of the inter-electronic angle of escape. In addition, we find that the previously observed  ``finger-like" structure in the correlated momenta of the strongly driven He is also present
 for strongly driven diatomic molecules. We show that it is mainly due to the strong interaction, backscattering, the recolliding electron undergoes from the remaining core.
  
\end{abstract}


 \maketitle 

\section{Introduction} 
 Over the last few years significant progress has been achieved in understanding the mechanisms underlying the correlated electron motion
 in the double ionization of atomic systems when driven by intense laser pulses of a few cycle duration and infrared frequencies.
Namely, for laser pulse intensities $10^{13}$-$10^{15}$ W/cm$^2$---non-sequential double ionization---the correlated electron dynamics is particularly pronounced and is understood in terms of the rescattering mechanism \cite{Corkum}. The latter is a three-step process where first, one electron tunnels to the continuum, then it is accelerated and, finally, it is driven back by the laser field to its parent ion
 where it transfers energy and liberates the still bound electron. It is now accepted that two are the main mechanisms leading to double 
 ionization in atomic systems: a) the rescattering electron upon its return to the nucleus transfers energy to the remaining electron with both electrons ejected simultaneously to the continuum--we refer to this mechanism as EI; b) the rescattering electron is ionized upon its return to the nucleus while exciting the remaining electron which is 
 ionized at a later time at a maximum of the laser pulse. The latter is the so-called recollision-induced excitation with subsequent field ionization mechanism (RESI)  \cite{RESI1, RESI2, RESI3,RESI4}.
 
 The correlated electron dynamics in the double ionization of strongly-driven diatomic molecules is a highly complex problem and is far from being theoretically understood. Responding to the two-center nuclear attraction, the correlated electron motion in the non-sequential double ionization  regime is much more complex than the atomic case with new phenomena emerging such as bond-softening \cite{bondsoftening} and alignment dependence---observed in molecular experiments \cite{align1, align2, align3}. 
The rescattering mechanism seems to be at the heart of the correlated electron dynamics for the case of two-electron diatomic molecules as was the case for the well studied He atom. 
In addition to EI and RESI, for the case of diatomic molecules an additional pathway is present---not previously identified in the double ionization of He. Namely, upon the return of the rescattering electron to the nucleus the two electrons form
 a doubly excited compound with both electrons ionizing at a latter time \cite{classicaleasy, strongBecker}---we refer to this 
 mechanism as the DE pathway.

 Given the state of the art in computational capabilities  addressing the double ionization of strongly driven systems 
with three-dimensional (3-d) first-principle techniques, namely quantum mechanical ones, is an immense task. Currently, 3-d quantum mechanical calculations from first principles (ab-initio) are available for the driven He atom \cite{strongHe}. To cope with the highly complex task of tackling the double ionization of diatomic molecules many studies use numerical quantum approaches of reduced dimensionality \cite{Andre1}. Others use judiciously chosen
 quantum mechanical models of reduced dimensionality, such as the one in ref. \cite{strongBecker}. Yet others use semi-analytical quantum approaches in the framework of the so-called Strong-Field Approximation \cite{strongfield}, not fully accounting for the Coulomb singularity.

 In the 
current paper, we study the correlated electron dynamics in the double ionization of diatomic molecules with ``frozen" nuclei in the non-sequential regime as a function of the intensity of the laser pulse. We do so, using a 3-d quasiclassical technique that we have first developed for conservative systems, namely, the multiple ionization of atomic systems such as Li, Be, by single photon absorption \cite{Agapi2}. We have very recently extended this technique to non-conservative systems treating the correlated electron dynamics of the He atom when driven by strong laser fields \cite{agapistrong} and attosecond pulses \cite{agapiatto}. Here we further build
on this technique by tackling more than one atomic centers.
The advantage of this technique is that it is numerically very efficient.  In addition, the method treats the Coulomb singularity with no approximation in contrast to techniques that use ``soft-core" potentials. Fully accounting for the Coulomb singularity with no approximations is important 
for describing accurately effects such as the striking so-called ``finger-like" structure which was recently observed \cite{Corkum1, Ullrich1}. This structure was attributed
to the strong interaction of the rescattering electron with the core---backscattering. 
In addition, accounting for the Coulomb singularity will be very important in pump-probe set-ups where VUV
or XUV pulses are used to probe the process. For these latter high frequency pulses, the excursion parameter of the electronic motion is smaller than the atomic dimensions making it very important to incorporate effects of the Coulomb potential in an exact manner.     

In section II of the current paper we first briefly present the 3-d quasiclassical model we use and we then study
the correlated momenta of the doubly ionized $N_{2}$ diatomic molecule as a function of the laser field intensity. We identify the three different
double ionization pathways for diatomic homonuclear molecules and show that each pathway leaves a distinct trace on the double differential probabilities of the emitted electrons. 
Finally, we show that the striking ``figure-like" structure in the correlated momenta identified first for atomic systems \cite{Corkum1,Ullrich1} is also present in the double ionization of diatomic molecules.

  \section{Quasiclassical Model} 
 Our 3-d quasiclassical model entails the following steps: We first set-up the initial phase space distribution of the two ``active" electrons in the N$_{2}$
 diatomic molecule. For intensities of the laser field that correspond to the tunneling regime we assume that one electron  
  tunnels
through the field-lowered Coulomb potential. In analogy to using the so-called ADK quantum mechanical rate formula for atoms \cite{tunnelingrateatom}
one can use quantum mechanical or semiclassical tunneling rate formulas for diatomic molecules, such as those described in refs. \cite{tunnelingratemolecule, semiclassical}. In the current work we use the rate provided in ref. \cite{semiclassical}. The longitudinal momentum is zero
while the transverse one is provided by a Gaussian distribution \cite{classicaldifficult}. The above description is valid when the instantaneous laser field 
is smaller than a threshold value which for the $N_{2}$ molecule with first ionization energy of $I_{p1}=0.5728$ a.u., second ionization energy $I_{p2}=0.9989$ a.u. and internuclear distance of 2.079 a.u. is around 0.075 a.u. To account for the field strengths corresponding to the over the barrier regime the initial phase space distribution of the two electrons is modeled by a double electron microcanonical distribution \cite{Olson}.

Next, we transform to a new system of coordinates, the so called ``regularized" coordinates \cite{regularized}. This transformation is exact
 and explicitly eliminates the Coulomb singularity.  This step is more challenging for molecular systems since one has to ``regularize" with respect to more than one atomic centers versus one atomic center for atoms.  Finally, we use Classical Trajectory Monte Carlo (CTMC) for the time propagation  \cite{CTMC}. The propagation involves the ``full" four-body Hamiltonian in the laser field with ``frozen" nuclei: $H=p_{1}^{2}/2+p_{2}^{2}/2
-1/|\vec{R/2}-\vec{r}_{1}|-1/|\vec{R/2}-\vec{r}_{2}|+1/|\vec{r}_{1}-\vec{r}_{2}|+(\vec{r}_{1}+\vec{r}_{2}) \cdot \vec{E}(t)$, with E(t) the electric field,
which is a cos pulse that is switched off with a $cos^{2}$ envelope, and $\vec{R}$ is the internuclear distance.

\section{Double Ionization Mechanisms}
\begin{figure}
\scalebox{0.27}{\includegraphics{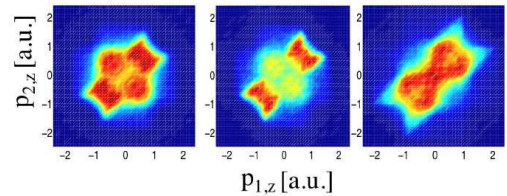}}
\caption{\label{fig1:pz1pz2all} The correlated electron momenta of the two escaping electrons parallel to the polarization axis. From left to right, we consider three laser pulses of 10$^{14}$ $Watts/cm^{2}$ and a 6-cycle duration, 10$^{14}$ $Watts/cm^{2}$ and a 3-cycle duration, and 1.5$\times$10$^{14}$ $Watts/cm^{2}$ and a 3-cycle duration.}  
\end{figure}

\begin{figure}
\scalebox{0.3}{\includegraphics{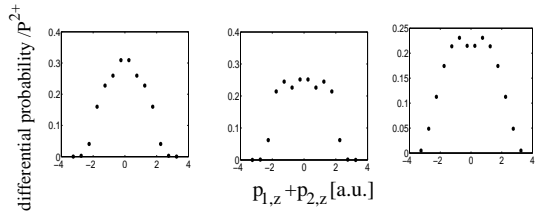}}
\caption{\label{fig2:mom} The sum of the parallel momentum distribution of the two escaping electrons. From left to right, we consider three laser pulses of 10$^{14}$ $Watts/cm^{2}$ and a 6-cycle duration, 10$^{14}$ $Watts/cm^{2}$ and a 3-cycle duration, and 1.5$\times$10$^{14}$ $Watts/cm^{2}$ and a 3-cycle duration.}
\end{figure}

We explore the double ionization of N$_{2}$ in the tunneling regime for a laser pulse intensity of 10$^{14}$ $Watts/cm^{2}$ and 1.5 10$^{14}$ $Watts/cm^{2}$, respectively. For the smaller intensity we consider
two pulses of 6 and 3-cycle duration. Our results are obtained for laser field polarization parallel to the molecular axis. We note that the initial state described in the previous section was shown
to describe very well double to single ratios of ionization \cite{classicaldifficult} and accurately capture the dependence of such ratios on the field polarization with respect to the molecular
axis. Future work will address including additional quantum mechanical effects into the model in order to accurately describe the dependence of, for instance, the correlated momenta on the
angle of the polarization axis with respect to the molecular axis. 
 From the correlated momenta of the two escaping electrons
shown in \fig{fig1:pz1pz2all} we confirm that for increasing duration of the laser pulse the probability for two electrons to get ionized through the EI pathway decreases. Indeed, the double hump structure in the sum of the parallel momenta distribution is much less pronounced for a long
laser pulse versus a short one, see \fig{fig2:mom}, as has already been pointed out in ref \cite{classicaleasy}. As for the case of the strongly driven helium, the maxima in the parallel momentum distribution correspond to 4$\sqrt{U_{p}}$,
with 2$\sqrt{U_{p}}$ the maximum velocity an electron can acquire from its interaction with the field. $U_{p}=E_{0}^{2}/(4\omega^{2})$ is the ponderomotive energy. Given the above discussion,
 we choose laser pulses of only 3-cycle duration in order for the traces of the EI pathway to be more pronounced and not be smeared out by the remaining mechanisms. 

\begin{figure}
\scalebox{0.3}{\includegraphics{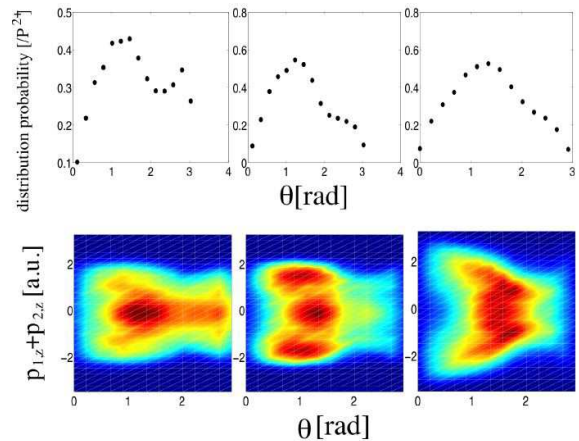}}
\caption{\label{fig3:pzangle} From left to right, we consider three laser pulses of 10$^{14}$ $Watts/cm^{2}$ and a 6-cycle duration, 10$^{14}$ $Watts/cm^{2}$ and a 3-cycle duration, and 1.5$\times$10$^{14}$ $Watts/cm^{2}$ and a 3-cycle duration.  We plot: a) (top panel) the distribution of the inter-electronic angles of escape binned in 14 intervals, 180$^{\circ}(l-1)/14<\theta<180^{\circ}/14\times l$ with l=1,...,14;
 b) (bottom panel) 
the sum of the parallel momenta as a function of the inter-electronic angle of escape.}
\end{figure}

\begin{figure}
\scalebox{0.3}{\includegraphics{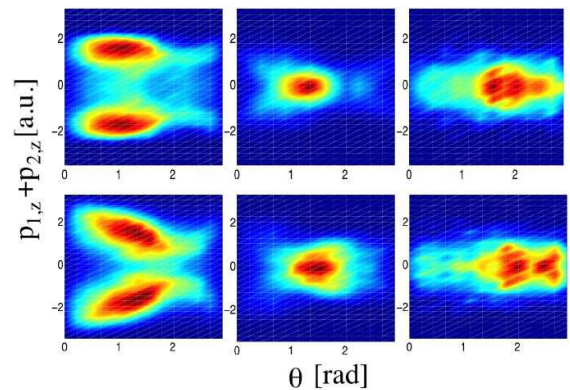}}
\caption{\label{fig4:pz1pz2mechanisms} Top panel corresponds to a laser pulse of 10$^{14}$ $Watts/cm^{2}$ and a 3-cycle duration; Bottom panel corresponds to a laser pulse of 10$^{14}$ $Watts/cm^{2}$ and a 3-cycle duration. From left to right we show the differential probability for the EI, RESI and DE mechanisms respectively.}
\end{figure}

The question we address first is whether the three different ionization mechanisms leave distinct traces when considering differential probabilities. In ref. \cite{strongBecker}, a planar quantum mechanical model is used with the motion of the center of mass of the two electrons restricted along the polarization axis. By monitoring in time the electron density as a function of two coordinates, that of the relative motion and that of the center of mass of the electron motion, it was shown that the ionization mechanisms leave distinct traces on the electron density. Our goal is to identify the 
different ionization mechanisms by their distinct fingerprints on the double differential probability of the sum of the parallel momenta as a function of the inter-electronic angle of escape, see \fig{fig3:pzangle}. To this end, we identify the recollision time (the time of minimum approach of the two electrons) through the maximum in the electron pair potential energy. Next, we identify the ionization time of the two electrons \cite{ionization}.  We then select three groups of trajectories. First, those where both electrons ionize close to the recollision time, which is close to (2/3+n)T with T the period of the field and label these trajectories as EI. Then, we select the trajectories where the recolliding electron ionizes close to the recollision time, while the remaining electron ionizes at a subsequent time close to a field maximum, and label them as RESI. Finally the trajectories where both electrons are ionized much later than the recollision time we label as DE. 
The double differential probability of the sum of the parallel momenta as a function of the inter-electronic angle is indeed different for each of the thus selected groups of trajectories, see \fig{fig4:pz1pz2mechanisms}. 

In particular, when the electrons ionize through the EI pathway, they escape with large momenta and with inter-electronic angles less than 90$^{\circ}$; for the trajectories that the electrons ionize through the RESI pathway the sum of the parallel components of the momenta is almost zero---consistent with our experience from strongly driven He---and escape with angles larger than those for the EI mechanism but yet mainly less than 90$^{\circ}$. Finally, when the electrons escape through the DE pathway the sum of the momenta parallel to the field is small, as for RESI, but the inter-electronic angle of escape is larger than 90$^{\circ}$. Given the above, it is clear from \fig{fig3:pzangle} that for an intensity of 10$^{14}$ $Watts/cm^{2}$ (tunneling regime) the longer the laser pulse the more pronounced is the contribution of the RESI and DE ionization pathways. Further more, for short laser pulses 
and intensities in the tunneling regime for larger intensities the momenta of the two
electrons are overall smaller, compare 10$^{14}$ versus 1.5$\times$ 10$^{14}$ $Watts/cm^{2}$ in \fig{fig4:pz1pz2mechanisms}.
 This is consistent, with our finding (not shown here) that the average recollision time shifts to smaller values for the larger intensity, which corresponds to the electrons having smaller momenta at the time of their release. Finally, comparing the sum of the momenta as a function of the inter-electronic angle for the strongly driven He, see ref. \cite{agapistrong}, with the molecular case shown in \fig{fig3:pzangle}, it is clear that the RESI is more pronounced for the molecular case, while the DE is not present for the atomic case.

\begin{figure}
\scalebox{0.27}{\includegraphics{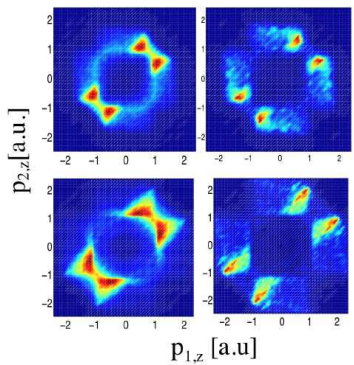}}
\caption{\label{fig5:back}Top panel corresponds to a laser pulse of 10$^{14}$ $Watts/cm^{2}$ and a 3-cycle duration; Bottom panel corresponds to a laser pulse of 10$^{14}$ $Watts/cm^{2}$ and a 3-cycle duration. The left figures are plotted using only the trajectories where $p_{1}\vee p_{2}>2\sqrt{U_{p}}$; The right figures are the same as the left ones except that in addition electron 2 is backscattering with $155^{\circ}<\vec{p}_{2,aft}\cdot \vec{p}_{2,bef}/|p_{1}p_{2}|<180^{\circ}$.  }
\end{figure}

In the case of the He atom, a striking ``finger-like" structure in the correlated momenta was very recently observed \cite{Corkum1, Ullrich1} and different aspects of it were discussed in refs. \cite{recoil, agapistrong}. Next, we investigate whether a ``finger-like" structure is also present in the molecular case and how it depends on the intensity of the laser pulse. To this end 
 we select those trajectories for which electron 2 (recolliding electron) backscatters from the nucleus, inverting the direction of its velocity. That is, $155^{\circ}<\vec{p}_{2,aft}\cdot \vec{p}_{2,bef}/|p_{2,aft}p_{2,bef}|<180^{\circ}$, with $\vec{p}_{2,bef/aft}$ the momentum of electron 2 just before and after the recollision time. The correlated momenta of the thus selected trajectories, as can be seen in \fig{fig5:back}, indeed correspond to a finger-like structure for the molecular case. As for the atomic case, we find that this structure extends beyond the 2$\sqrt{U_{p}}$ maximum momentum limit---the maximum momentum that can be acquired from the field.  Note first
that this structure persists for both 10$^{14}$ and 1.5$\times$ 10$^{14}$  $Watts/cm^{2}$. In somewhat more details in \fig{fig5:back} (left panel) we show the structure for the correlated momenta with at least one of the two momenta having magnitude greater than $2\sqrt{U_{p}}$. We note that the trajectories shown in \fig{fig5:back} (right panel) are a subset of those in \fig{fig5:back} (left panel) and that for the remaining trajectories either 
electron 2 or electron 1 reverses its velocity but with a smaller recoil angle, that is, $90^{\circ}<\vec{p}_{i,aft}\cdot \vec{p}_{i,bef}/|p_{i,aft}p_{i,bef}|<150^{\circ}$.
It is worth noting that we obtain the finger-like structure in \fig{fig5:back} in the first and third quadrant for electrons
escaping asymptotically with an angle around 50$^\circ$ which is larger than for the He case where for the recoil trajectories the two electrons escape almost parallel to each other.

\begin{figure}
\scalebox{0.27}{\includegraphics{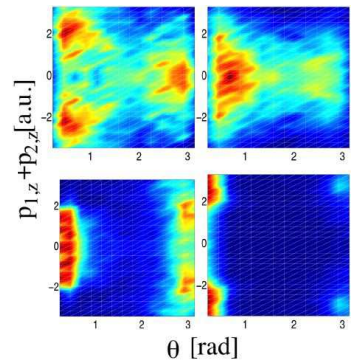}}
\caption{\label{fig6:high}Double differential probability for a laser pulse of 10$^{15}$ $Watts/cm^{2}$ and a 3-cycle duration; 
Top panel corresponds to the trajectories that ionize over the barrier with those on the left corresponding to the two electrons ionizing with a time difference less than half a period and those
on the right ionizing with a time difference larger than half a period. There is no maximum in the electron-electron potential energy for these trajectories and thus no recollision time.  
Bottom panel corresponds to the trajectories that ionize below the barrier (tunneling) with those on the left corresponding to electrons ionizing with small time difference and close to the recollision time; those on the right correspond to electrons ionizing with larger time difference with the recolliding electron ionizing close to the recollision time. For these trajectories the maximum of the potential energy shifts to a maximum of the field and thus the recollision time is smaller compared to the recollision time at 10$^{14}$ and 1.5 10$^{14}$ $Watts/cm^{2}$.  }
\end{figure}

The distinct traces of the ionization pathways on double differential probabilities and the ``finger-like" structure are present for intensities of the laser pulse in the tunneling regime. Considering
a high intensity of 10$^{15}$ $Watts/cm^{2}$ (over the barrier regime) we find that the ``finger-like" structure in the first and third quadrant disappears. For this high intensity we have trajectories 
ionizing when the instantaneous value of the field of the laser pulse is below the barrier and trajectories that ionize when the strength of the field is above the barrier.  For those in the tunneling regime,
we can still define a recollision time from the maximum of the potential energy with the difference that the recollision time shifts
to earlier times at a maximum of the field compared to the recollision time of the smaller intensities previously considered. Thus, when the electrons are released close to the recollision time their energy of release is small in contrast to the previously studied intensities.
This is consistent with \fig{fig6:high} where in the bottom panel we see that when the electrons ionize with small time difference and close to the recollision time  they escape with smaller momenta compared to those the electrons
have for smaller intensities (EI pathway). It can also explain that the electrons ionizing with larger time difference with the recolliding electron ionizing close to the recollision time 
escape with higher momenta than those for smaller intensities (RESI mechanism);  the remaining electron is not released at a maximum of the field resulting in higher momentum. We also note that the inter-electronic angle of escape is overall smaller
in this case due to the large strength of the field which pulls the electrons along its direction. Finally, electrons ionizing when the instantaneous field is above the barrier
 have larger momenta when the time difference in their ionization time is small, see top panel in \fig{fig6:high}. In this latter case there is no maximum in the potential energy and thus
we can not define a recollision time.

In conclusion, using a 3-d quassicalssical method we explore the different mechanisms for double ionization of a strongly driven homonuclear diatomic molecule. We find that the ionization
mechanisms leave distinct traces on the sum of the momenta parallel to the field polarization of the escaping electrons as a function of the inter-electronic angle of escape for intensities in the
tunneling regime. In addition, the previously observed in atomic systems ``finger-like" structure is present for molecular systems as well and disappears (in the first and the third quadrant)
for high intensities in the over the barrier regime. 
Future studies will aim to incorporate in the current model interference effects from the presence of the two nuclei and explore 
the ``fingerprints" of interference on the correlated momenta of the escaping electrons.

 \end{document}